\pdfoutput=1
\documentclass[11pt]{article}

\usepackage[final]{acl}
\usepackage{fancyhdr}

\usepackage{times}
\usepackage{latexsym}

\usepackage[T1]{fontenc}
\usepackage[utf8]{inputenc}

\usepackage{microtype}

\usepackage{inconsolata}

\usepackage{graphicx}
\usepackage{amsmath}
\usepackage{multirow}
\usepackage{booktabs}
\usepackage{rotating}
\usepackage{subcaption}

\title{Evaluating Cognitive-Behavioral Fixation \\via Multimodal User Viewing Patterns on Social Media}

\author{
    \textbf{Yujie Wang\textsuperscript{1,2}},
    \textbf{Yunwei Zhao\textsuperscript{3}},
    \textbf{Jing Yang\textsuperscript{1,2}},
    \textbf{Han Han\textsuperscript{3}},
    \textbf{Shiguang Shan\textsuperscript{1,2}},
    \textbf{Jie Zhang\textsuperscript{1,2}}
    \\
    \\
    \textsuperscript{1}State Key Laboratory of AI Safety, Institute of Computing Technology, \\Chinese Academy of Sciences
    \\
    \textsuperscript{2}University of Chinese Academy of Sciences
    \\
    \textsuperscript{3}CNCERT/CC
    \\
    \small{
        \textbf{Correspondence:} \href{mailto:zhangjie@ict.ac.cn}{zhangjie@ict.ac.cn}
    }
}

\begin{document}
    \maketitle

    \thispagestyle{fancy}
    \fancyhf{} % 清空默认的页眉页脚
    \fancyfoot[C]{\small This paper has been accepted for presentation at the EMNLP 2025 main conference.}
    \renewcommand{\headrulewidth}{0pt} % 去掉页眉横线
    \renewcommand{\footrulewidth}{0pt} % 去掉页脚横线

    \begin{abstract}
        Digital social media platforms frequently contribute to cognitive-behavioral fixation, a phenomenon in which users exhibit sustained and repetitive engagement with narrow content domains.
        While cognitive-behavioral fixation has been extensively studied in psychology, methods for computationally detecting and evaluating such fixation remain underexplored.
        To address this gap, we propose a novel framework for assessing cognitive-behavioral fixation by analyzing users' multimodal social media engagement patterns.
        Specifically, we introduce a multimodal topic extraction module and a cognitive-behavioral fixation quantification module that collaboratively enable adaptive, hierarchical, and interpretable assessment of user behavior.
        Experiments on existing benchmarks and a newly curated multimodal dataset demonstrate the effectiveness of our approach, laying the groundwork for scalable computational analysis of cognitive fixation.
        All code in this project is publicly available for research purposes at \url{https://github.com/Liskie/cognitive-fixation-evaluation}
    \end{abstract}

    \section{Introduction}\label{sec:introduction}

    Digital media and online platforms significantly shape user behavior through personalized algorithms and constant connectivity, which curate the information users encounter and structure their engagement patterns.
    This pervasive mediation often amplifies existing preferences and habits, potentially leading to self-reinforcing cognitive and behavioral feedback loops~\cite{delvicario2016spreading, cinelli2021echo}.
    In other words, rather than merely reflecting individual preferences, social media and recommendation systems may actively shape and constrain perceptions, preferences, and decision-making processes.

    While such tailored content enhances user experience, it also raises concerns about \textbf{cognitive-behavioral fixation} phenomena, which is defined as obsessive preoccupations with specific ideas or activities, impairing balanced information processing and flexible thinking~\cite{dielenberg2024biological}.
    Common manifestations include misinformation loops~\cite{delvicario2016spreading}, echo chambers~\cite{cinelli2021echo}, and compulsive engagement patterns~\cite{markett2023social}.
    For example, misinformation loops emerge when recommendation algorithms continually amplify users' preexisting misconceptions with ideologically aligned but factually inaccurate content ~\cite{delvicario2016spreading}.
    Similarly, echo chambers restrict users to homogeneous viewpoints, exacerbating confirmation bias and social polarization~\cite{cinelli2021echo}. Meanwhile, compulsive engagement driven by persuasive design (e.g., endless feeds, personalized notifications) fosters addictive and repetitive usage behaviors~\cite{markett2023social}.
    Collectively, these fixation effects distort perceptions of reality, entrench false beliefs, deepen societal divisions, and negatively impact mental health, with consequences including anxiety, diminished well-being, and reduced attention spans~\cite{markett2023social}.

    Addressing cognitive-behavioral fixation is crucial not only for safeguarding individual mental health but also for maintaining societal stability, as fixation-driven behaviors (e.g., widespread misinformation acceptance, extreme polarization) can undermine public discourse and trust in information ecosystems.
    Despite these risks, a significant research gap remains: \textbf{no existing computational framework can automatically detect and quantify cognitive-behavioral fixation within online environments}.
    Existing studies often treat echo chambers, misinformation spread, and compulsive behaviors separately, lacking a unified model to quantify fixation holistically.
    Psychological studies, while insightful, primarily rely on qualitative methods such as case studies, surveys, or controlled experiments~\cite{meloy2020cognitive}, making them difficult to scale for automated, real-time digital behavior analysis.

    In response, we formalize the novel task of computational cognitive-behavioral fixation evaluation for social media users.
    We propose a multimodal analytical framework designed to automatically detect and quantify fixation based on users' digital interactions, specifically focusing on text posts and video content.
    Our framework supports adaptive, hierarchical, and interpretable evaluation of cognitive-behavioral fixation in real-world social online environments.

    \begin{figure*}[th]
        \centering
        \includegraphics[width=0.7\textwidth]{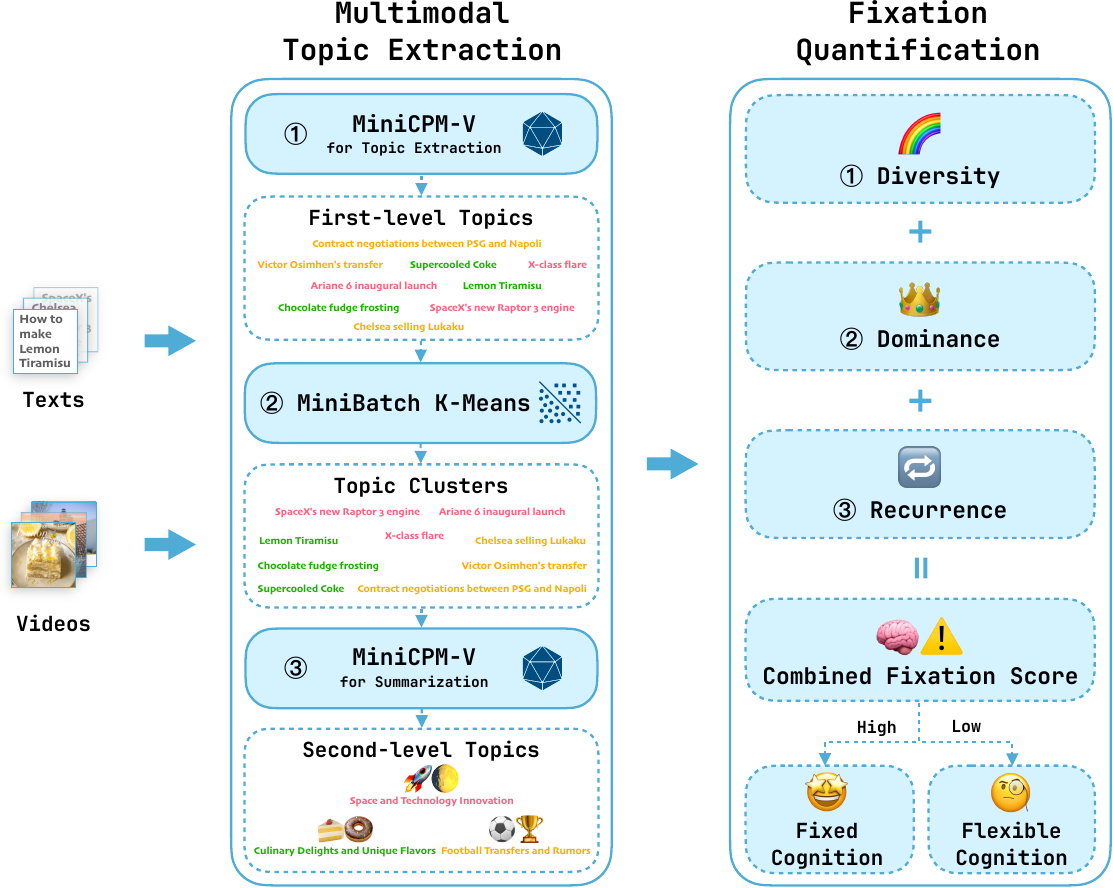}
        \caption{Overview of our proposed framework for cognitive-behavioral fixation evaluation.}
        \label{fig:overview}
    \end{figure*}

    The key contributions of this work are:

    \begin{enumerate}
        \item We introduce the first formalization of cognitive-behavioral fixation evaluation as a computational task for social media behavior analysis.
        \item We propose an \textbf{adaptive}, \textbf{hierarchical}, and \textbf{interpretable} framework that generalizes across modalities and dataset scales, extracts multi-level user interests, and quantifies fixation.
    \end{enumerate}

    \section{Related Work}\label{sec:related-work}

    \subsection{Cognitive-Behavioral Fixation in Psychology}\label{subsec:cognitive-behavioral-fixation-in-psychology}
    Cognitive-behavioral fixation describes obsessive preoccupation with specific ideas or behaviors that impairs flexible thinking and decision-making~\cite{dielenberg2024biological}.
    Psychology traditionally studies fixation through clinical observation and case studies, linking it to confirmation bias, belief perseverance, and cognitive rigidity~\cite{lord1979biased, meloy2020cognitive}.
    Such mechanisms clearly manifest in online contexts like echo chambers and misinformation loops, where repetitive exposure reinforces narrow beliefs and resistance to counter-evidence, exacerbating polarization and impairing public discourse~\cite{zollo2017debunking, cinelli2021echo}.
    While psychology provides rich theoretical foundations, it lacks automated, scalable methods for detecting fixation in large-scale digital behavior data—a gap this work addresses.

    \subsection{Topic Extraction}
    Classical topic modeling approaches, such as Latent Dirichlet Allocation (LDA)~\cite{blei2001latent} and Non-negative Matrix Factorization (NMF)~\cite{lee1999learning}, enable unsupervised extraction of themes but struggle with short or noisy texts.
    Recent neural topic models, including Embedded Topic Model (ETM) and Contextualized Topic Models (CTM), leverage embedding-based representations to enhance coherence and scalability~\cite{bianchi2021pre, dieng2020topic}.
    DeTiME further improved coherence through diffusion-enhanced large language model embeddings~\cite{xu-etal-2023-detime}.
    Multimodal topic extraction methods, such as Multimodal LDA~\cite{zhu2010mmlda} and CLIP-based models~\cite{luo2022clip4clip}, integrate visual and textual inputs into a unified semantic space.
    Our work uniquely applies multimodal topic extraction for cognitive-level behavioral analysis.

    \subsection{Behavioral Analysis and Metrics}
    Quantitative metrics such as Shannon entropy, KL divergence, and burstiness have widely analyzed user engagement to characterize topical diversity, concentration, and temporal patterns.
    Shannon entropy measures topical diversity, where lower entropy reflects narrower user focus~\cite{weng2012competition, sonoda2022analyzing, song2014evaluating}.
    Topic concentration is typically quantified using the Herfindahl–Hirschman Index (HHI)~\cite{zhu2015topicality}, and KL divergence captures shifts in sequential user interests~\cite{sritrakool2021personalized}.
    Burstiness metrics based on inter-event timing highlight clustered activity patterns indicating compulsive or reactionary behavior~\cite{goh2008burstiness, karsai2018bursty}.
    Entropy-based methods have also quantified ideological fixation and selective exposure~\cite{munoz2024quantifying, pratelli2024entropy}.
    These existing metrics form the foundation for our multimodal fixation evaluation framework.

    \section{Methodology}

    \subsection{Problem Formulation}

    We focus on analyzing users' multimodal browsing behaviors on social media platforms to quantitatively evaluate cognitive-behavioral fixation.

    Formally, for each user \( u \), we define their viewing history as an ordered sequence:
    \begin{equation}
        \mathcal{H}^{(u)} = \{h_1, h_2, \dots, h_T\}
    \end{equation}
    where each interaction \( h_t \) corresponds to a viewed content item (e.g., a post or video) at timestamp \( t \).
    Each \( h_t \) contains both textual and visual information.
    From each content \( h_t \), we extract a set of fine-grained topic tags:
    \begin{equation}
        T(h_t) = \{\text{topic}_{t,1}, \dots, \text{topic}_{t,m}\}
    \end{equation}
    representing the semantic essence of the content.
    These topics form the basis for later analyses.

    Given the temporal sequence of extracted topics across all viewed content, our goal is to assess the extent to which the user's attention is narrowly concentrated, repetitive, and persistent over time, i.e., key signatures of cognitive-behavioral fixation.

    \subsection{Overall Framework}

    Our framework enables \textbf{adaptive}, \textbf{hierarchical}, and \textbf{interpretable} analysis of cognitive-behavioral fixation in real-world social media environments.

    \paragraph{Multimodal Hierarchical Topic Extraction}
    This module processes user interactions with textual and visual content to extract two levels of topic representations.
    The first-level consists of fine-grained topic phrases that summarize individual content items.
    These are subsequently grouped into second-level topic phrases that represent broader thematic categories.
    These hierarchical topic phrases captures both immediate and abstract semantic interests of users.

    \paragraph{Cognitive-Behavioral Fixation Quantification}
    Using the generated topic phrases, this module computes behavioral metrics (i.e., diversity, dominance, and recurrence) that reflect key dimensions of cognitive-behavioral fixation.
    These are integrated into a unified, interpretable fixation score that quantifies the intensity and persistence of users’ topical engagement.
    The framework’s adaptability to diverse datasets and scales ensures its applicability across various social media environments.

    \subsection{Multimodal Hierarchical Topic Extraction Module}\label{subsec:multimodal-hierarchical-topic-extraction-module}

    \subsubsection{Extraction of First-level Topics}

    The primary objective of first-level topic extraction is to identify concise semantic summaries for each piece of multimodal content, effectively capturing users' immediate content interests at a granular level.
    These first-level labels serve as a foundation for subsequent higher-level semantic analysis.

\begin{figure}[t]
        \centering
        \includegraphics[width=0.45\textwidth]{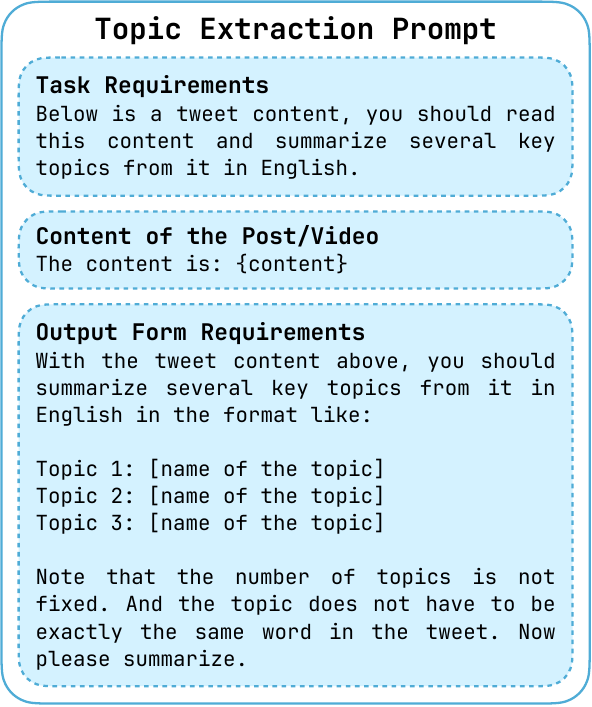}
        \caption{Prompt for extracting topics from multimodal content using MiniCPM-V.}
        \label{fig:topic_extraction_prompt}
    \end{figure}

    To achieve this, we employ the MiniCPM-V model to generate a set of short topic phrases from each video's textual description and visual frames, summarizing the core themes of the content.
    These phrases constitute the \textbf{first-level labels}.
    The prompt used for topic extraction via MiniCPM-V is illustrated in Figure~\ref{fig:topic_extraction_prompt}.

    \subsubsection{Extraction of Second-level Topics}

    Second-level topic extraction aims to cluster semantically related first-level topics into broader thematic categories.
    This higher-level grouping facilitates interpretable and scalable analysis of users' overall content interests, enabling the identification of thematic patterns indicative of cognitive-behavioral fixation.

    Specifically, we follow these steps. Each first-level topic phrase is encoded using the SentenceBERT~\cite{reimers-gurevych-2019-sentence} model to obtain dense semantic embeddings.
    These embeddings are then grouped into \(K\) cohesive clusters using the MiniBatch K-means algorithm.
    We select \(K\) by a sweep over $\{100,200,300,400\}$ using intra-/inter-cluster distance and their ratio, and adopt $K{=}300$ as a fidelity–interpretability trade-off (see Table~\ref{tab:k_sweep} in Appendix~\ref{sec:implementation-details}).
    Each resulting cluster represents a distinct \textbf{second-level semantic topic class}, encapsulating broader thematic areas.

\begin{figure}[t]
        \centering
        \includegraphics[width=0.45\textwidth]{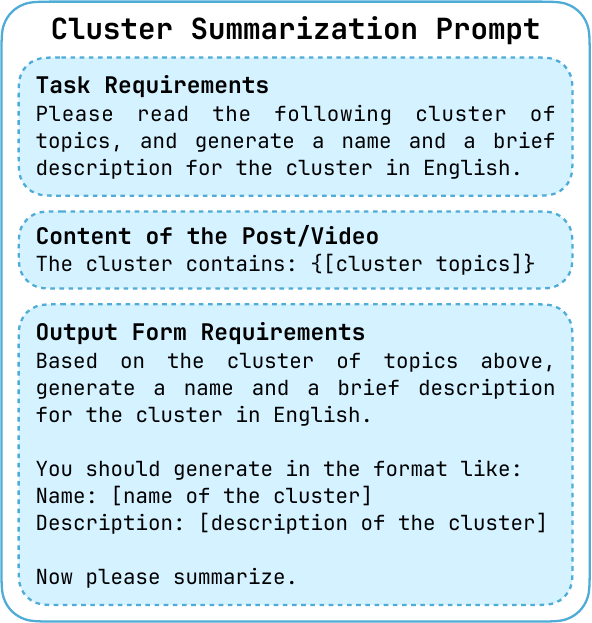}
        \caption{Prompt for summarizing topic clusters into human-readable labels.}
        \label{fig:cluster_summarization_prompt}
    \end{figure}

    To ensure interpretability, we generate descriptive, interpretable names for each cluster.
    A representative sample (e.g., 100 topic phrases) is randomly selected from each cluster.
    MiniCPM-V is then used to summarize these sampled phrases into concise, descriptive labels, as shown in Figure~\ref{fig:cluster_summarization_prompt}.

    Each content item is consequently associated with second-level topic labels, reflecting the thematic clusters of its first-level topics.

    \subsection{Cognitive-Behavioral Fixation Quantification Module}\label{subsec:cognitive-behavioral-fixation-quantification}

    We propose a unified metric to evaluate cognitive-behavioral fixation by integrating three core behavioral dimensions: \textbf{Diversity}, \textbf{Dominance}, and \textbf{Recurrence}.
    While the individual metrics we employ, i.e., Shannon entropy, Herfindahl-Hirschman Index (HHI), and burstiness, are well-established in prior work, our primary contribution lies in their combination into a cohesive, interpretable, and scalable framework specifically designed to assess fixation behaviors in multimodal social media contexts.

    \textbf{Diversity} is captured using Shannon entropy~\cite{shannon1948mathematical}, which measures the unpredictability in the distribution of topic clusters.
    For a sliding window of \( w \) days ending at time \( t \), let \( p_k \) denote the proportion of user interactions assigned to topic cluster \( k \) (out of \( K \) total clusters):
    $\bar{H}_t = -\sum_{k=1}^{K} p_k \log p_k$.
    To enable cross-user comparability, we normalize this value:
    \begin{equation}
        \bar{H}_t^{\text{norm}} = \frac{\bar{H}_t}{\log(K)}
    \end{equation}

    \textbf{Dominance} is measured using the Herfindahl--Hirschman Index (HHI)~\cite{herfindahl1950concentration, hirschman1945national}, which quantifies the concentration of user attention across topics:
    \begin{equation}
        \bar{D}_t^{\text{HHI}} = \sum_{k=1}^{K} p_k^2
    \end{equation}

    \textbf{Recurrence} is assessed via burstiness~\cite{goh2008burstiness, karsai2018bursty}, which reflects how clustered in time the re-engagements with the same topic are.
    For each topic cluster, we compute the inter-event intervals \(\tau\) between successive user interactions, and define:
    \begin{equation}
        \bar{R}_t^{\text{burst}} = \frac{\sigma_\tau - \mu_\tau}{\sigma_\tau + \mu_\tau}
    \end{equation}
    where \(\mu_\tau\) and \(\sigma_\tau\) are the mean and standard deviation of inter-event intervals.
    The final recurrence score is aggregated across all clusters.

    We then apply MinMax normalization to each component and combine them to compute a unified fixation score.
    \begin{equation}
        \bar{F}_t^{(w)} = \alpha \cdot (1 - \bar{H}_t^{\text{norm}}) + \beta \cdot \bar{D}_t^{\text{HHI}} + \gamma \cdot (1 - \bar{R}_t^{\text{burst}})
        \label{eq:fix}
    \end{equation}
    where \(\alpha\), \(\beta\), and \(\gamma\) are tunable hyperparameters controlling the relative influence of each behavioral dimension.
    A higher fixation score indicates stronger cognitive-behavioral fixation tendencies.

    By unifying these metrics, we enable interpretable evaluation of user fixation behavior that can be scaled across large multimodal datasets.

    \section{Experimental Setup}

    \subsection{Datasets}

    We evaluate our multimodal topic extraction methods on three datasets: a widely-used benchmark for text-based topic modeling, and two multimodal datasets, including a new dataset we have constructed for cognitive-behavioral analysis in social media browsing contexts.
    We preprocess each dataset following previous works\cite{grootendorst2022bertopic, bianchi-etal-2021-cross}.
    Brief data statistics are listed below.

    \paragraph{20 Newsgroups}~\cite{Lang95} contains 18,000 English newsgroup posts distributed across 20 categories, commonly used for text topic modeling\footnote{\url{http://qwone.com/~jason/20Newsgroups/}}.
    It enables the evaluation of both topic diversity and coherence.

    \paragraph{COCO 2017}~\cite{gonzalez2024neural} composes of 120,000 images labeled with objects (80+ categories) and five captions per image\footnote{\url{http://cocodataset.org/}}.
    We adopt it for evaluating multimodal topic modeling, following prior work .

    \paragraph{X User Browsing Dataset (XUB, ours)} is a newly collected multimodal dataset comprising browsing histories of 163 anonymized users on platform X, spanning a two-month period.
    Each record includes textual content, visual features, and timestamps.
    The dataset contains a total of 1,169,041 records, averaging 7,172 entries per user, and covers 212,289 unique posts.
    XUB enables the evaluation of cognitive-behavioral fixation and topic engagement dynamics over time.

    \subsection{Baseline Methods}

    We compare our method against five baselines.

    For text-only models, we include LDA~\cite{blei2001latent}, a foundational probabilistic topic model; BERTopic~\cite{grootendorst2022bertopic}, which integrates embeddings and class-based TF-IDF; and QualIT~\cite{kapoor2024qualitative}, which iteratively refines topic quality.
    For multimodal methods, we include Multimodal-ZeroShotTM~\cite{gonzalez2024neural}, which jointly decodes text and image features, and M3L-Contrast~\cite{zosa-pivovarova-2022-multilingual}, which applies contrastive learning for robust cross-modal alignment.

    \subsection{Evaluation Metrics}

    \paragraph{Topic Diversity.}
    Measures how well the extracted topics span distinct semantic areas, computed as the proportion of unique tokens among the top-\(k\) keywords across topics.
    Higher scores indicate broader coverage and less redundancy.

    \paragraph{Topic Coherence.}
    Assesses the semantic consistency of keywords within each topic using \(C_v\).
    Higher values suggest stronger internal cohesion and better interpretability.

    These two metrics are applied to all datasets to evaluate topic quality.

    \paragraph{Cognitive-Behavioral Fixation Score.}
    On the XUB dataset, we use our composite fixation score based on diversity, dominance, and recurrence, as introduced in Section~\ref{subsec:cognitive-behavioral-fixation-quantification}, and also report each component metric for detailed analysis.

    \begin{table*}[htb]
        \centering
        \resizebox{\textwidth}{!}{
            \begin{tabular}{cl|cc}
                \toprule
                \textbf{Dataset} & \textbf{Model}                                        & \textbf{Topic Coherence} & \textbf{Topic Diversity} \\
                \hline\hline
                \multirow{4}{*}{\textbf{20ng}}
                & LDA~\cite{blei2001latent}                             & 47.0\%                   & 69.0\%                   \\
                & BERTTopic~\cite{grootendorst2022bertopic}             & 56.0\%                   & 82.0\%                   \\
                & QualIT~\cite{kapoor2024qualitative}                   & \textbf{66.0\%}          & \underline{95.0\%}       \\
                & \textbf{Ours (Text-only)}                             & \underline{62.3\%}       & \textbf{96.5\%}          \\
                \hline\hline
                \multirow{3}{*}{\textbf{COCO}}
                & Multimodal-ZeroShotTM~\cite{bianchi-etal-2021-cross}  & 54.0\%                   & \underline{60.0\%}       \\
                & M3L-Contrast~\cite{zosa-pivovarova-2022-multilingual} & \underline{56.0\%}       & 47.0\%                   \\
                & \textbf{Ours (Multimodal)}                            & \textbf{75.0\%}          & \textbf{80.5\%}          \\
                \hline\hline
                \multirow{7}{*}{\textbf{XUB}}
                & LDA~\cite{blei2001latent}                             & 24.2\%                   & 59.4\%                   \\
                & BERTTopic~\cite{grootendorst2022bertopic}             & 44.5\%                   & 72.7\%                   \\
                & Multimodal-ZeroShotTM~\cite{gonzalez2024neural}       & 63.4\%                   & 68.0\%                   \\
                & M3L-Contrast~\cite{zosa-pivovarova-2022-multilingual} & 74.7\%                   & 20.8\%                   \\
                \cline{2-4}
                & \textbf{Ours (Text-only)}                             & \textbf{77.8}\%          & \underline{75.9}\%       \\
                & \textbf{Ours (Video-only)}                            & 75.5\%                   & 69.7\%                   \\
                & \textbf{Ours (Multimodal)}                            & \underline{75.6\%}       & \textbf{88.2\%}          \\
                \bottomrule
            \end{tabular}
        }
        \caption{Topic Coherence and Topic Diversity for different methods. The best results are bolded, and the second-best results are underlined.}
        \label{tab:topic_baseline}
    \end{table*}

    \subsection{Human Annotation \& Validation Setup}
    To empirically ground our fixation score, we designed a human annotation and validation protocol on a subset of XUB. We sampled 30 anonymized users and prepared, for each user, (i) full interaction logs (timestamps and topic-cluster assignments) and (ii) a topic word cloud for rapid sense-making. Three trained annotators independently assigned a user-level label: \emph{fixated} vs.\ \emph{not fixated}. We used majority vote as gold labels and measured inter-annotator agreement with Fleiss' $\kappa$.

    For validation, we split the 30-user gold set using 3-fold stratified 10-repeat cross-validation.
    In each run, we: (a) compute each metric (diversity, dominance, recurrence) and the composite fixation score on the training fold; (b) select the decision threshold for the composite using Youden’s $J$ statistic; (c) report performance (Accuracy, Precision, Recall, F1) on the held-out fold for ablations (Diversity, Dominance, Recurrency; pairwise combinations; and full combination).
    We preregistered the protocol to report per-fold means and standard deviations.
    The full instructions for annotators are provided in Appendix~\ref{sec:human-annotation}.

    \begin{figure*}[t]
        \centering
        \begin{subfigure}[b]{0.45\textwidth}
            \includegraphics[width=\linewidth]{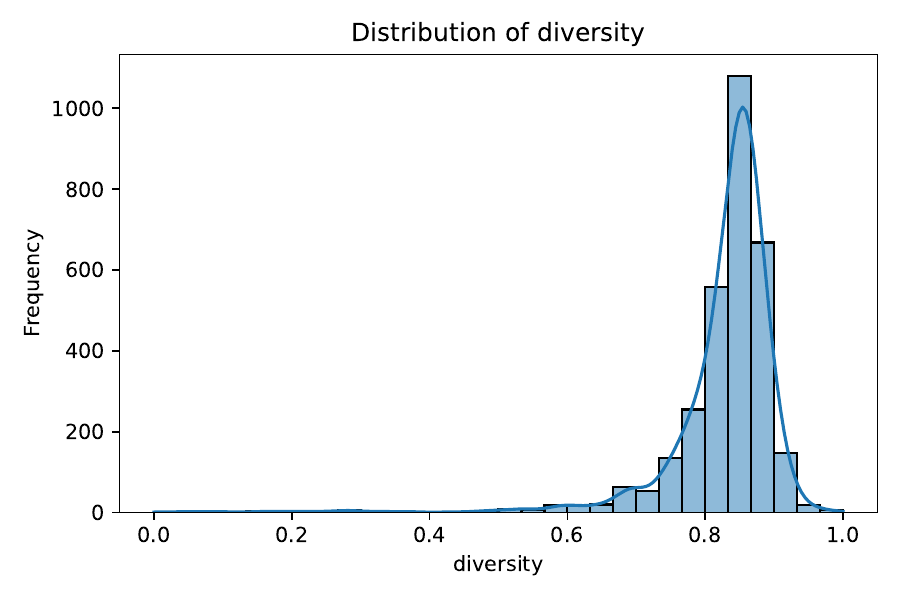}
            \caption{Diversity}
            \label{fig:subfig1}
        \end{subfigure}
        \hfill
        \begin{subfigure}[b]{0.45\textwidth}
            \includegraphics[width=\linewidth]{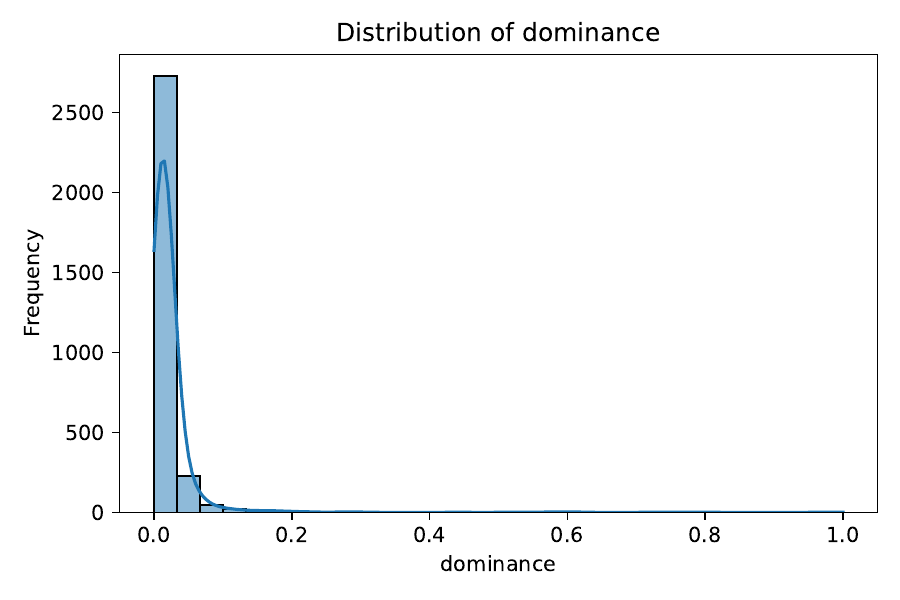}
            \caption{Dominance}
            \label{fig:subfig2}
        \end{subfigure}
        \vskip\baselineskip
        \begin{subfigure}[b]{0.45\textwidth}
            \includegraphics[width=\linewidth]{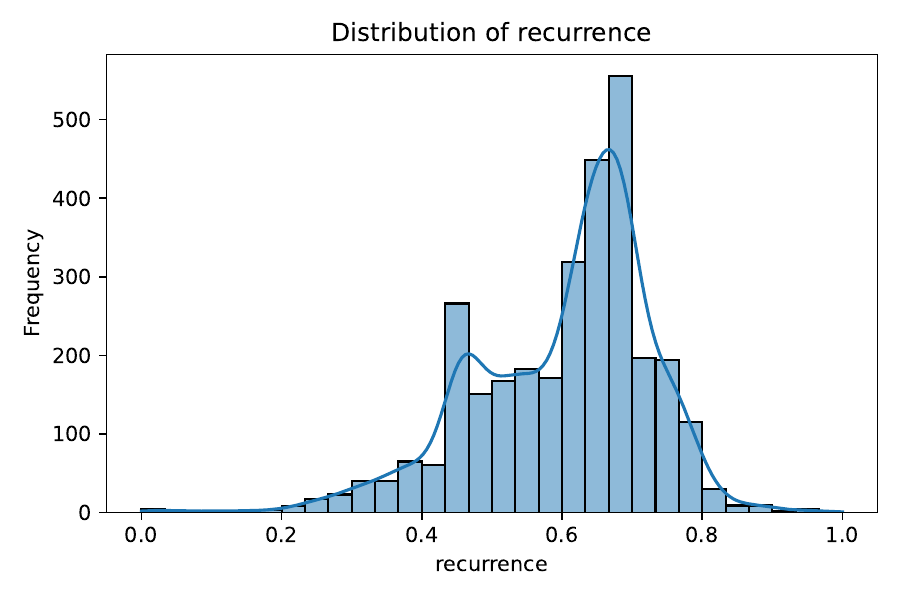}
            \caption{Recurrence}
            \label{fig:subfig3}
        \end{subfigure}
        \hfill
        \begin{subfigure}[b]{0.45\textwidth}
            \includegraphics[width=\linewidth]{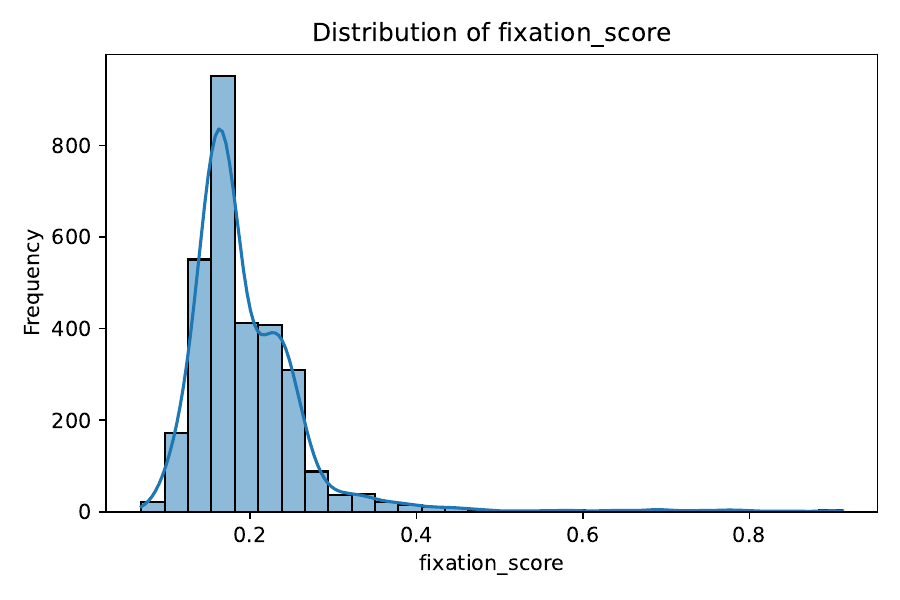}
            \caption{Combined Score}
            \label{fig:subfig4}
        \end{subfigure}
        \caption{The distribution of average fixation metrics of each user on our XUB dataset.}
        \label{fig:2x2_grid}
    \end{figure*}

    \section{Results and Analysis}

    \subsection{Topic Extraction Performance}

    We evaluated our multimodal topic extraction method on the 20 Newsgroups, COCO 2017, and our XUB datasets.
    As shown in Table~\ref{tab:topic_baseline}, the results indicate that in the 20 Newsgroups dataset, our method achieved a topic coherence score of 62.3\%, second after QualIT (66.0\%), but excelled in topic diversity with a score of 96.5\%.
    The slightly lower topic coherence compared to QualIT's may be due to QualIT's specialized refinement techniques for text-based topics.

    For the COCO dataset, which includes both text and visual content, our method achieved a topic coherence score of 75.0\% and a diversity score of 80.5\%, outperforming other methods in both metrics.
    These results validate our motivation to use multimodal data to improve topic extraction, showing that our approach effectively integrates textual and visual information to produce coherent and diverse topics.
    On the XUB dataset, our method demonstrated strong robustness across different settings, aligning with our goal of building a flexible and adaptive topic modeling framework suitable for analyzing complex social media content.
    However, we observed that topic coherence was lower in the multimodal setting compared to the text-only baseline.
    This decline may be attributed to the presence of visual elements in the images that are weakly related to the textual themes, i.e., visual noise, which can introduce semantically less relevant keywords into the model, ultimately reducing overall topic coherence.
    The performance of the clustering procedure is deferred to Appendix~\ref{sec:topic-clustering-performance}

    The results affirm the efficacy of our VLM-based approach in achieving high topic coherence and diversity, especially in multimodal settings.
    Combining visual and textual data leads to more comprehensive and interpretable topic models, making our method well-suited for complex, real-world applications like social media analysis.

%    \begin{figure*}[t]
%        \centering
%        \includegraphics[width=1\textwidth]{figures/user_case}
%        \caption{The trends of fixation metrics of users in XUB dataset.}
%        \label{fig:fixation_score_trends}
%    \end{figure*}

    \begin{figure}[th]
        \centering
        \begin{subfigure}[b]{0.8\linewidth}
            \includegraphics[width=\linewidth]{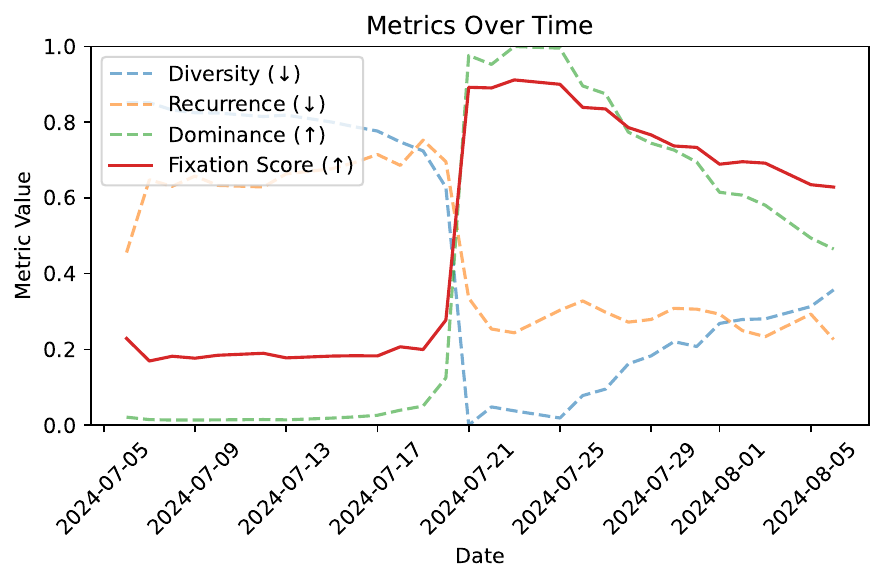}
            \caption{User \#1}
            \label{fig:metrics_1}
        \end{subfigure}
        \vskip\baselineskip
        \begin{subfigure}[b]{0.8\linewidth}
            \includegraphics[width=\linewidth]{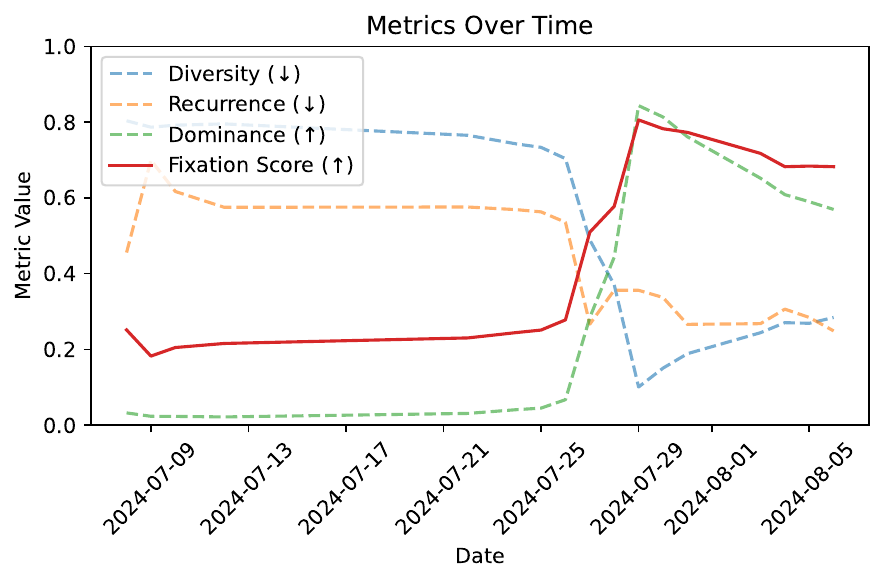}
            \caption{User \#2}
            \label{fig:metrics_2}
        \end{subfigure}
        \vskip\baselineskip
        \begin{subfigure}[b]{0.8\linewidth}
            \includegraphics[width=\linewidth]{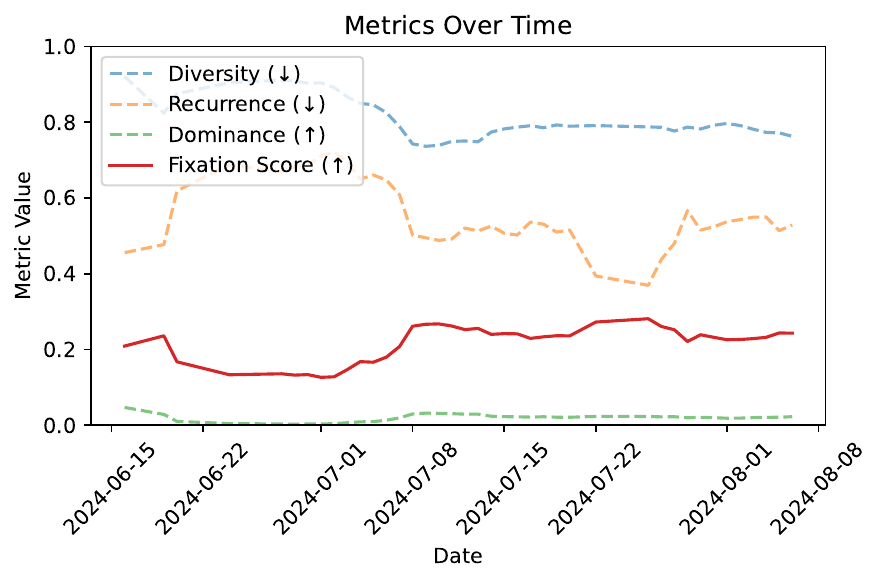}
            \caption{User \#3}
            \label{fig:metrics_3}
        \end{subfigure}
        \caption{The trends of fixation metrics of users in XUB dataset.}
        \label{fig:fixation_score_trends}
    \end{figure}

    \begin{figure}[th]
        \centering
        \begin{subfigure}[b]{0.8\linewidth}
            \includegraphics[width=\linewidth]{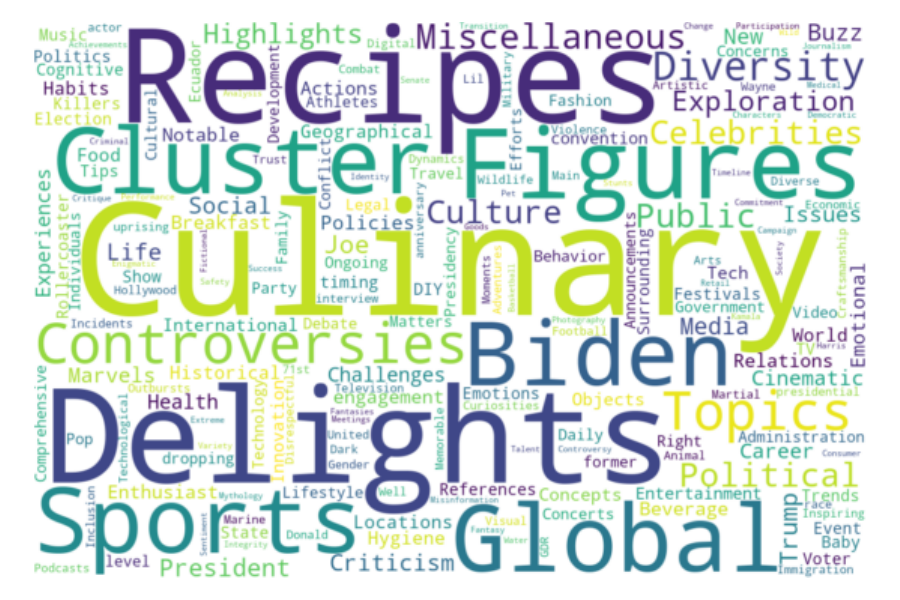}
            \caption{User \#1}
            \label{fig:wordcloud_1}
        \end{subfigure}
        \vskip\baselineskip
        \begin{subfigure}[b]{0.8\linewidth}
            \includegraphics[width=\linewidth]{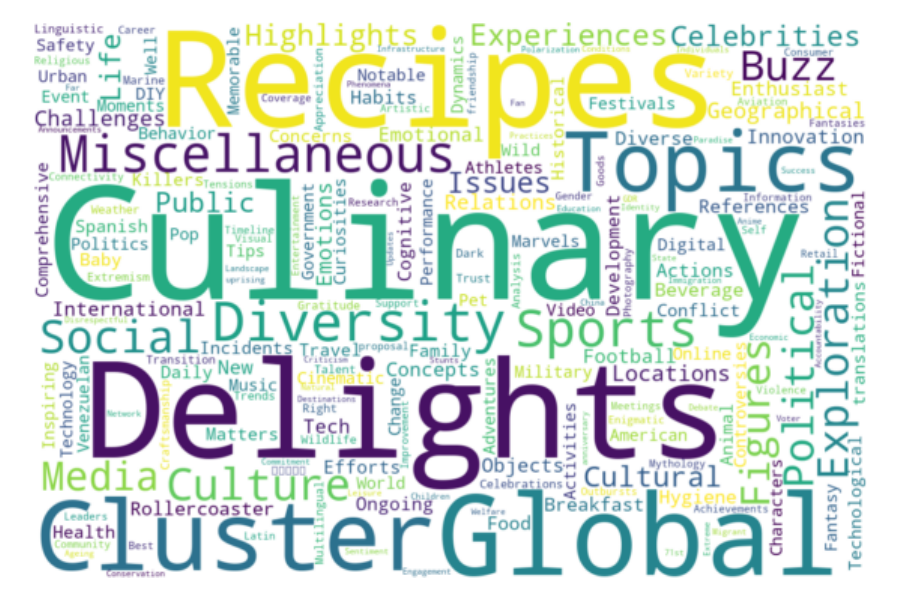}
            \caption{User \#2}
            \label{fig:wordcloud_2}
        \end{subfigure}
        \vskip\baselineskip
        \begin{subfigure}[b]{0.8\linewidth}
            \includegraphics[width=\linewidth]{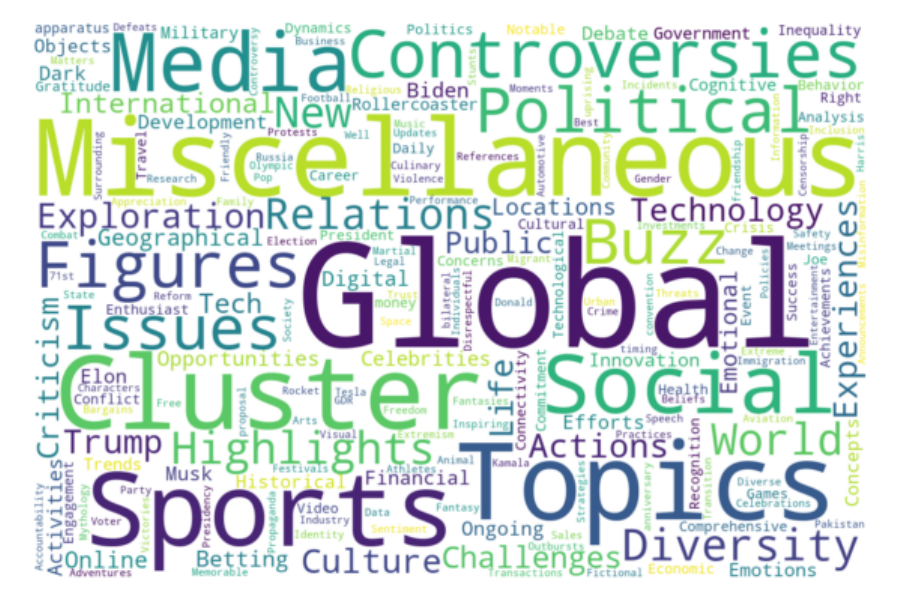}
            \caption{User \#3}
            \label{fig:wordcloud_3}
        \end{subfigure}
        \caption{The word cloud of users in XUB dataset.}
        \label{fig:wordclouds}
    \end{figure}

    \subsection{Cognitive-Behavioral Fixation Evaluation Results}
    In this study, we utilized a set of metrics to assess cognitive-behavioral fixation, including diversity, dominance, recurrence, and the combined fixation score.
    These metrics provide a multidimensional perspective on how users engage with topics, revealing patterns that may indicate fixation.

    As shown in Figure~\ref{fig:2x2_grid}, the diversity distribution is concentrated between 0.6 and 0.95, with a peak around 0.87.
    This indicates that participants generally exhibit a moderate level of topic diversity.
    The relatively high diversity values suggest that while users do not exclusively focus on a single topic, their engagement is not highly diverse, hinting at a potential fixation on a limited set of topics.
    According to the dominance, while there is no extreme dominance of a single topic, we can still observe that the top topic could be 3 to 4 time more focused that the tail topics.
    This dominance level reflects a balanced but slightly focused cognitive-behavioral pattern, indicating a potential fixation tendency without complete exclusivity.
    The recurrence distribution peaks near 0.65.
    Recurrence is measured using burstiness, where lower values indicate more regular and temporally concentrated re-engagement with specific topics.
    This pattern suggests a relatively strong cognitive-behavioral recurrence, which is a key characteristic of fixation behavior.

    The fixation score's ability to capture fixation trends demonstrates its interpretability not only conceptually but also empirically.
    The distribution of users' combined fixation scores is centered around 0.2, extending to 0.5 at most.
    Using the cross-validated cut-off ($\geq 0.352$), we identify strong fixation tendencies; scores below $\sim0.2$ suggest a more exploratory pattern.
    This is demonstrated in Section~\ref{subsec:case-studies} detailedly.
    The fixation score captures the overall fixation pattern in real-world data, offering a more comprehensive assessment of cognitive-behavioral fixation than any single metric alone.

    \subsection{Human Annotation \& Validation Results}
    \label{subsec:human-validation-results}
    Annotator agreement was substantial (Fleiss’ $\kappa=0.524$). Cross-validated threshold selection yielded an optimal fixation-score cut-off of $0.352\pm0.007$ (Youden’s $J$), which we use in all analyses below.

    \paragraph{Ablation and performance.}
    We assessed each component metric and their combinations against the gold labels.
    Dominance alone achieved the strongest single-metric F1, diversity was competitive, while recurrence alone was weaker but contributed when combined.
    The full composite score achieved accuracy $0.857\pm0.094$.
    Detailed results are shown in Table~\ref{tab:human-ablations}.

    \begin{table*}[t]
        \centering
        \begin{tabular}{lccccc}
            \toprule
            \textbf{Metrics} & \textbf{$\tau$}    & \textbf{Accuracy}            & \textbf{Precision}           & \textbf{Recall} & \textbf{F1} \\
            \midrule
            Diversity        & 0.309\textpm{}0.069          & 0.830\textpm{}0.144          & 0.816\textpm{}0.282          & 0.650\textpm{}0.288 & 0.667\textpm{}0.223 \\
            Dominance        & 0.053\textpm{}0.016          & 0.830\textpm{}0.156          & 0.797\textpm{}0.288          & 0.672\textpm{}0.289 & 0.679\textpm{}0.233 \\
            Recurrence       & 0.654\textpm{}0.038          & 0.700\textpm{}0.139          & 0.506\textpm{}0.371          & 0.428\textpm{}0.279 & 0.407\textpm{}0.228 \\
            \midrule
            Div.+Dom.        & 0.182\textpm{}0.043          & 0.830\textpm{}0.144          & 0.816\textpm{}0.282          & 0.650\textpm{}0.288 & 0.667\textpm{}0.223 \\
            Div.+Rec.        & 0.499\textpm{}0.002          & 0.857\textpm{}0.094          & 0.850\textpm{}0.248          & 0.589\textpm{}0.258 & 0.667\textpm{}0.219 \\
            Dom.+Rec.        & 0.348\textpm{}0.007          & 0.860\textpm{}0.089          & 0.781\textpm{}0.321          & 0.589\textpm{}0.324 & 0.638\textpm{}0.281 \\
            \midrule
            \textbf{All}     & \textbf{0.352\textpm{}0.007} & \textbf{0.857\textpm{}0.094} & \textbf{0.850\textpm{}0.248} & \textbf{0.589\textpm{}0.258} & \textbf{0.667\textpm{}0.219} \\
            \bottomrule
        \end{tabular}
        \caption{Cross-validated results on the 30-user gold set (mean$\pm$std).}
        \label{tab:human-ablations}
    \end{table*}

    \paragraph{User-level analysis.}
    Among the 163 users in XUB, 14 users (8.59\%) exceed the $0.352$ threshold, spanning domains such as sports (5), cooking (4), games (2), politics (1), global affairs (1), and technology (1).
    We observe long fixation streaks (up to 24 consecutive days) and gradual disengagement patterns (e.g., cooking-focused users declining from $\sim0.91$ to $\sim0.6$).
    Overall, fixation durations ranged from 7 to 24 consecutive days in the 30-day window.

    \subsection{Case Studies}\label{subsec:case-studies}
    In this section, we first visualized the changes in the fixation scores of three users over time, as shown in Figure~\ref{fig:fixation_score_trends}.
    It can be observed that user \#1 had a relatively low fixation score before July 17, 2024, indicating that the user was exposed to a diverse range of video topics during that period.
    This suggests that the recommendation platform had not yet accurately identified the user's preferences, reflecting an exploratory phase of interest.
    As time passed, the user's fixation score increased slightly.
    This implies that the platform may have begun pushing more targeted content based on the user's click and dwell behaviors.
    On July 21, 2024, user \#1's fixation score showed a sharp increase, accompanied by a significant drop in diversity and a rise in dominance.
    This trend suggests a decrease in the diversity of content the user was exposed to and an increase in the concentration of dominant topics, pointing to more focused recommendation results and a clear convergence of user preferences.
    Subsequently, the fixation score remained at a relatively high level, indicating that the user's content consumption behavior had become strongly fixated.
    This may suggest that the user had entered the cognitive-behavioral fixation state, i.e., being repeatedly exposed to homogeneous content, potentially limiting their cognitive scope and the diversity of perspectives they encounter.

    Then, we analyze the second-level topics in user \#1's viewing history.
    The most frequently occurring categories were Class 14, 123, 0, 48, and 246, corresponding to topics such as Culinary Delights and Recipes, Culinary Exploration, Celebrities and Public Figures, Cinematic Marvel, and Culinary Diversity, respectively.
    The word cloud in Figure~\ref{fig:wordcloud_1} further illustrates user \#1's video content preferences, with prominent keywords such as Culinary, Delights, Biden, and Recipes.
    These high-frequency topics indicate that the user's video consumption was primarily concentrated in the areas of cooking and politics, showing a clear focus in their interests.
    This aligns with the trend observed in fixation scores and reinforces the idea that the user's cognitive interest narrowed over time.

    User \#2 followed a similar pattern, with the fixation score rising over time alongside decreasing diversity and increasing dominance.
    This indicates a shift toward more repetitive and focused content consumption, suggesting emerging cognitive-behavioral fixation.

    In contrast, User \#3 exhibited a different pattern, with fixation scores remaining relatively stable over time and no significant shifts in diversity or dominance. The associated word cloud also reveals a broad range of topics without a clear focus, suggesting more varied content consumption and an absence of strong cognitive-behavioral fixation.

    \section{Conclusion}\label{sec:conclusion}
    In this work, we present a novel framework for computationally assessing cognitive-behavioral fixation by analyzing users’ multimodal engagement on social media.
    Our approach integrates an \textbf{adaptive} topic extraction process, a \textbf{hierarchical} representation of user interests via multi-level topic phrases, and an \textbf{interpretable} fixation quantification module based on diversity, dominance, and recurrence.
    Experiments on existing benchmarks and a newly constructed multimodal dataset demonstrate the effectiveness of our method, establishing a foundation for large-scale, automated analysis of fixation behaviors.
    Future directions include broadening modality inclusion, refining temporal granularity, and incorporating causal inference techniques to deepen understanding of fixation’s formation mechanisms and societal impacts in digital ecosystems.

    \section*{Limitations}

    While our proposed framework provides an adaptive, hierarchical, and interpretable approach to evaluating cognitive-behavioral fixation in multimodal social media environments, several limitations remain.

    \paragraph{Limited Modalities.}
    Although we incorporate both textual and visual modalities, our analysis does not currently include audio or interaction-based features (e.g., click sequences, comments, dwell time), which may provide additional behavioral signals relevant to fixation.

    \paragraph{Model Dependency.}
    Our topic extraction depends on the performance and bias of the MiniCPM-V vision-language model and SentenceBERT. Limitations in these models, such as cultural bias or reduced performance in underrepresented domains, may influence the accuracy and generalizability of the extracted topics.

    \paragraph{Composite-weight selection.}
    Our unified fixation score (Eq.~\ref{eq:fix}) currently uses equal weights for diversity, dominance, and recurrence.
    While neutral and interpretable, different applications may warrant data-driven or stakeholder-informed weights; learning or calibrating them against behavioral outcomes or human judgments is a valuable direction.

    \paragraph{Absence of demographics and external generalization.}
    XUB is anonymized and lacks demographic attributes (e.g., age, gender), limiting demographic correlation analyses.
    Moreover, we are not aware of public datasets with timestamped multimodal browsing logs suitable for external validation; future work will explore collaborations to evaluate generalization across populations and platforms.

    \paragraph{Unmodeled cues and visual noise.}
    We focus on cognitive–behavioral signals and do not incorporate affective polarity, semantic drift, or engagement entropy.
    These cues may enrich interpretation and will be explored.
    In addition, visual content can introduce semantic noise that slightly degrades coherence; visual-feature denoising or saliency filtering is a promising extension.

    \section*{Ethical Considerations}
    In conducting this research, we have adhered to ethical standards and have taken steps to ensure that no new ethical concerns were introduced.

    \paragraph{Data Usage.} We have fully adhered to the data usage policies of all data sources.
    Privacy and confidentiality are paramount, and any data used in our study has been anonymized to prevent the identification of individual users.

    \paragraph{Code and Transparency.} The source code for baseline models used in our study are either open-sourced or licensed for academic purposes.
    We are committed to transparency in our research, ensuring that all results and methodologies are thoroughly documented and accessible for replication and scrutiny by the research community.

    \paragraph{PII Anonymization in Our Dataset.} For our newly collected XUB dataset, we have ensured that all personally identifiable information (PII), including usernames, profile images, and timestamps that could lead to re-identification, have been removed or anonymized.
    The dataset only retains behavioral features necessary for fixation analysis and cannot be traced back to individual users.
    The users are all volunteer students in our lab, no profit is made from the data, and all users have given their consent for their data to be used in this research.

    \section*{Acknowledgements}
    This work is partially supported by the Strategic Priority Research Program of the Chinese Academy of Sciences (No. XDB0680202), the Key Research and Development Program of Xinjiang Uyghur Autonomous Region (No. 2024B03026), Beijing Nova Program (No. 20230484368) and Youth Innovation Promotion Association, CAS.

    \bibliography{custom}

\begin{thebibliography}{33}
\providecommand{\natexlab}[1]{#1}

\bibitem[{Bianchi et~al.(2021{\natexlab{a}})Bianchi, Terragni, and Hovy}]{bianchi2021pre}
Federico Bianchi, Silvia Terragni, and Dirk Hovy. 2021{\natexlab{a}}.
\newblock \href {https://doi.org/10.18653/v1/2021.acl-short.96} {Pre-training is a hot topic: Contextualized document embeddings improve topic coherence}.
\newblock In \emph{Proceedings of the 59th Annual Meeting of the Association for Computational Linguistics and the 11th International Joint Conference on Natural Language Processing (Volume 2: Short Papers)}, pages 759--766, Online. Association for Computational Linguistics.

\bibitem[{Bianchi et~al.(2021{\natexlab{b}})Bianchi, Terragni, Hovy, Nozza, and Fersini}]{bianchi-etal-2021-cross}
Federico Bianchi, Silvia Terragni, Dirk Hovy, Debora Nozza, and Elisabetta Fersini. 2021{\natexlab{b}}.
\newblock \href {https://doi.org/10.18653/v1/2021.eacl-main.143} {Cross-lingual contextualized topic models with zero-shot learning}.
\newblock In \emph{Proceedings of the 16th Conference of the European Chapter of the Association for Computational Linguistics: Main Volume}, pages 1676--1683, Online. Association for Computational Linguistics.

\bibitem[{Blei et~al.(2001)Blei, Ng, and Jordan}]{blei2001latent}
David~M. Blei, Andrew~Y. Ng, and Michael~I. Jordan. 2001.
\newblock \href {https://api.semanticscholar.org/CorpusID:272555142} {Latent dirichlet allocation}.
\newblock \emph{J. Mach. Learn. Res.}, 3:993--1022.

\bibitem[{Cinelli et~al.(2021)Cinelli, Morales, Galeazzi, Quattrociocchi, and Starnini}]{cinelli2021echo}
Matteo Cinelli, Gianmarco De~Francisci Morales, Alessandro Galeazzi, Walter Quattrociocchi, and Michele Starnini. 2021.
\newblock \href {https://doi.org/10.1073/pnas.2023301118} {The echo chamber effect on social media}.
\newblock \emph{Proceedings of the National Academy of Sciences}, 118(9):e2023301118.

\bibitem[{Dielenberg(2024)}]{dielenberg2024biological}
Robert~A. Dielenberg. 2024.
\newblock \href {https://www.academia.edu/124300425/The_biological_foundations_of_fixation_a_general_theory} {The biological foundations of fixation: a general theory}.
\newblock \emph{Academia Biology}, 3.

\bibitem[{Dieng et~al.(2020)Dieng, Wang, Gao, and Paisley}]{dieng2020topic}
Adji~B. Dieng, Chong Wang, Jianfeng Gao, and John Paisley. 2020.
\newblock Topic modeling in embedding spaces.
\newblock \emph{Transactions of the Association for Computational Linguistics}, 8:439--453.

\bibitem[{Goh and Barab{\'a}si(2008)}]{goh2008burstiness}
Kwang-Il Goh and Albert-L{\'a}szl{\'o} Barab{\'a}si. 2008.
\newblock \href {https://doi.org/10.1209/0295-5075/81/48002} {Burstiness and memory in complex systems}.
\newblock \emph{Europhysics Letters}, 81(4):48002.

\bibitem[{Gonz{\'a}lez-Pizarro and Carenini(2024)}]{gonzalez2024neural}
Felipe Gonz{\'a}lez-Pizarro and Giuseppe Carenini. 2024.
\newblock Neural multimodal topic modeling: A comprehensive evaluation.
\newblock \emph{arXiv preprint arXiv:2403.17308}.

\bibitem[{Grootendorst(2022)}]{grootendorst2022bertopic}
Maarten Grootendorst. 2022.
\newblock Bertopic: Neural topic modeling with a class-based tf-idf procedure.
\newblock \emph{arXiv preprint arXiv:2203.05794}.

\bibitem[{Herfindahl(1950)}]{herfindahl1950concentration}
Orris~Clemens Herfindahl. 1950.
\newblock \emph{Concentration in the US Steel Industry}.
\newblock Columbia University.

\bibitem[{Hirschman(1945)}]{hirschman1945national}
Albert~O Hirschman. 1945.
\newblock National power and the structure of foreign trade.
\newblock Technical report, University of California Press.

\bibitem[{Kapoor et~al.(2024)Kapoor, Gil, Bhaduri, Mittal, and Mulkar}]{kapoor2024qualitative}
Satya Kapoor, Alex Gil, Sreyoshi Bhaduri, Anshul Mittal, and Rutu Mulkar. 2024.
\newblock Qualitative insights tool (qualit): Llm enhanced topic modeling.
\newblock \emph{arXiv preprint arXiv:2409.15626}.

\bibitem[{Karsai et~al.(2018)Karsai, Jo, and Kaski}]{karsai2018bursty}
M{\'a}rton Karsai, Hang-Hyun Jo, and Kimmo Kaski. 2018.
\newblock \href {https://doi.org/10.1007/978-3-319-78511-7} {Bursty human dynamics}.
\newblock \emph{Springer Briefs in Complexity}.

\bibitem[{Lang(1995)}]{Lang95}
Ken Lang. 1995.
\newblock Newsweeder: Learning to filter netnews.
\newblock In \emph{Proceedings of the Twelfth International Conference on Machine Learning}, pages 331--339.

\bibitem[{Lee and Seung(1999)}]{lee1999learning}
Daniel~D. Lee and H.~Sebastian Seung. 1999.
\newblock Learning the parts of objects by non-negative matrix factorization.
\newblock \emph{Nature}, 401(6755):788--791.

\bibitem[{Lord et~al.(1979)Lord, Ross, and Lepper}]{lord1979biased}
Charles~G. Lord, Lee Ross, and Mark~R. Lepper. 1979.
\newblock Biased assimilation and attitude polarization: The effects of prior theories on subsequently considered evidence.
\newblock \emph{Journal of Personality and Social Psychology}, 37(11):2098--2109.

\bibitem[{Luo et~al.(2022)Luo, Wang, Zhang, and Yu}]{luo2022clip4clip}
Shikun Luo, Wen Wang, Yongchao Zhang, and Philip~S. Yu. 2022.
\newblock Clip4clip: An empirical study of clip for end-to-end video clip retrieval.
\newblock In \emph{Proceedings of the 30th ACM International Conference on Multimedia (MM)}, pages 1307--1316.

\bibitem[{Markett and Montag(2023)}]{markett2023social}
Silja Markett and Christian Montag. 2023.
\newblock \href {https://doi.org/10.3389/fpubh.2023.1123} {Social media use and everyday cognitive failure: investigating the role of fear of missing out and social network use disorder}.
\newblock \emph{Frontiers in Public Health}, 11:1123.

\bibitem[{Meloy and Rahman(2020)}]{meloy2020cognitive}
J.~Reid Meloy and Tahir Rahman. 2020.
\newblock \href {https://doi.org/10.1037/tam0000141} {The creation of america latina asociacion of threat assessment professionals}.
\newblock \emph{Journal of Threat Assessment and Management}, 7(3-4):111--126.

\bibitem[{Mu{\~n}oz et~al.(2024)Mu{\~n}oz, Samory, and Mitra}]{munoz2024quantifying}
Alberto Mu{\~n}oz, Mattia Samory, and Tanushree Mitra. 2024.
\newblock \href {https://doi.org/10.1140/epjds/s13688-024-00384-1} {Quantifying polarization in online political discourse}.
\newblock \emph{EPJ Data Science}, 13(1):5.

\bibitem[{Pratelli et~al.(2024)Pratelli, Del~Vicario, and Quattrociocchi}]{pratelli2024entropy}
Andrea Pratelli, Michela Del~Vicario, and Walter Quattrociocchi. 2024.
\newblock \href {https://doi.org/10.1093/pnasnexus/pgae051} {Entropy-based detection of twitter echo chambers}.
\newblock \emph{PNAS Nexus}, 3(1):pgae051.

\bibitem[{Reimers and Gurevych(2019)}]{reimers-gurevych-2019-sentence}
Nils Reimers and Iryna Gurevych. 2019.
\newblock \href {https://doi.org/10.18653/v1/D19-1410} {Sentence-{BERT}: Sentence embeddings using {S}iamese {BERT}-networks}.
\newblock In \emph{Proceedings of the 2019 Conference on Empirical Methods in Natural Language Processing and the 9th International Joint Conference on Natural Language Processing (EMNLP-IJCNLP)}, pages 3982--3992, Hong Kong, China. Association for Computational Linguistics.

\bibitem[{Shannon(1948)}]{shannon1948mathematical}
Claude~E Shannon. 1948.
\newblock A mathematical theory of communication.
\newblock \emph{Bell System Technical Journal}, 27(3):379--423.

\bibitem[{Song et~al.(2014)Song, Luo, He, Raghavan, Hsu, and Giles}]{song2014evaluating}
Yang Song, Hao Luo, Luo He, Anirudh Raghavan, Bo-June~Paul Hsu, and Lee Giles. 2014.
\newblock \href {https://doi.org/10.1145/2600428.2609625} {Evaluating and predicting user engagement change in web search}.
\newblock In \emph{Proceedings of the 37th International ACM SIGIR Conference on Research and Development in Information Retrieval}, pages 323--332.

\bibitem[{Sonoda et~al.(2022)Sonoda, Watanabe, and Ueda}]{sonoda2022analyzing}
Takuya Sonoda, Kohei Watanabe, and Michiko Ueda. 2022.
\newblock \href {https://doi.org/10.1007/s42001-021-00132-1} {Analyzing user engagement in news apps considering diversity}.
\newblock \emph{Journal of Computational Social Science}, 5:735--752.

\bibitem[{Sritrakool and Maneeroj(2021)}]{sritrakool2021personalized}
Nakarin Sritrakool and Saranya Maneeroj. 2021.
\newblock \href {https://doi.org/10.1109/DSAA53316.2021.9564194} {Personalized preference drift aware sequential recommender}.
\newblock In \emph{2021 IEEE 8th International Conference on Data Science and Advanced Analytics (DSAA)}, pages 1--10.

\bibitem[{Vicario et~al.(2016)Vicario, Bessi, Zollo, Petroni, Scala, Caldarelli, Stanley, and Quattrociocchi}]{delvicario2016spreading}
Michela~Del Vicario, Alessandro Bessi, Fabiana Zollo, Fabio Petroni, Antonio Scala, Guido Caldarelli, H.~Eugene Stanley, and Walter Quattrociocchi. 2016.
\newblock \href {https://doi.org/10.1073/pnas.1517441113} {The spreading of misinformation online}.
\newblock \emph{Proceedings of the National Academy of Sciences}, 113(3):554--559.

\bibitem[{Weng et~al.(2012)Weng, Flammini, Vespignani, and Menczer}]{weng2012competition}
Lilian Weng, Alessandro Flammini, Alessandro Vespignani, and Filippo Menczer. 2012.
\newblock \href {https://doi.org/10.1038/srep00335} {Competition among memes in a world with limited attention}.
\newblock \emph{Scientific Reports}, 2:335.

\bibitem[{Xu et~al.(2023)Xu, Hu, Wu, and Sengamedu}]{xu-etal-2023-detime}
Weijie Xu, Wenxiang Hu, Fanyou Wu, and Srinivasan Sengamedu. 2023.
\newblock \href {https://doi.org/10.18653/v1/2023.findings-emnlp.606} {{D}e{T}i{ME}: Diffusion-enhanced topic modeling using encoder-decoder based {LLM}}.
\newblock In \emph{Findings of the Association for Computational Linguistics: EMNLP 2023}, pages 9040--9057, Singapore. Association for Computational Linguistics.

\bibitem[{Zhu et~al.(2010)Zhu, Ahmed, and Xing}]{zhu2010mmlda}
Jun Zhu, Ahmed Ahmed, and Eric~P. Xing. 2010.
\newblock Medlda: Maximum margin supervised topic models.
\newblock \emph{Journal of Machine Learning Research}, 13:2237--2278.
\newblock Adaptable for multimodal extensions like mmLDA.

\bibitem[{Zhu et~al.(2015)Zhu, Sobhani, and Guo}]{zhu2015topicality}
Xiaodan Zhu, Parinaz Sobhani, and Hongyu Guo. 2015.
\newblock \href {https://doi.org/10.1371/journal.pone.0140555} {Topicality and impact in social media: Diverse messages, focused messengers}.
\newblock \emph{PLOS ONE}, 10(10):e0140555.

\bibitem[{Zollo et~al.(2017)Zollo, Bessi, Vicario, Scala, Caldarelli, Shekhtman, Havlin, and Quattrociocchi}]{zollo2017debunking}
Fabiana Zollo, Alessandro Bessi, Michela~Del Vicario, Antonio Scala, Guido Caldarelli, Louis Shekhtman, Shlomo Havlin, and Walter Quattrociocchi. 2017.
\newblock Debunking in a world of tribes.
\newblock \emph{Scientific Reports}, 7(1):1--9.

\bibitem[{Zosa and Pivovarova(2022)}]{zosa-pivovarova-2022-multilingual}
Elaine Zosa and Lidia Pivovarova. 2022.
\newblock \href {https://aclanthology.org/2022.coling-1.355/} {Multilingual and multimodal topic modelling with pretrained embeddings}.
\newblock In \emph{Proceedings of the 29th International Conference on Computational Linguistics}, pages 4037--4048, Gyeongju, Republic of Korea. International Committee on Computational Linguistics.

\end{thebibliography}

    \appendix

    \section{Implementation Details}\label{sec:implementation-details}

    For topic phrase generation, we employ the MiniCPM-V-2.6 vision-language model to process multimodal inputs comprising both textual and visual data.
    Semantic embeddings are obtained using the SentenceBERT~\cite{reimers-gurevych-2019-sentence} model to encode first-level topic phrases.
    These embeddings are clustered using the MiniBatch K-means algorithm.
    We set the number of second-level topic clusters (\(K\)) by balancing granularity and interpretability.
    A larger $K$ yields more specific themes but risks fragmenting semantically close topics; a smaller $K$ merges distinct themes and obscures nuance.
    We performed a quantitative sweep over $K\in\{100, 200, 300, 400\}$ on XUB using three standard clustering quality indicators computed in the SentenceBERT embedding space: (i) average intra-cluster distance (lower is better), (ii) average inter-cluster distance (higher is better), and (iii) their ratio (higher indicates better separation).
    As shown in Table~\ref{tab:k_sweep}, all indicators improved monotonically with larger $K$, but the marginal gain diminished beyond $K{=}300$.
    We therefore adopt $K{=}300$ as a practical compromise between fidelity and interpretability for all main experiments.

    \begin{table}[t]
        \centering
        \small
        \begin{tabular}{lccc}
            \toprule
            \textbf{$K$ (clusters)} & \textbf{Intra $\downarrow$} & \textbf{Inter $\uparrow$} & \textbf{Inter/Intra $\uparrow$} \\
            \midrule
            100                     & 1.1981                      & 0.6049                    & 0.505                           \\
            200                     & 1.1703                      & 0.6542                    & 0.559                           \\
            300                     & 1.1514                      & 0.6869                    & 0.597                           \\
            400                     & 1.1381                      & 0.7082                    & 0.622                           \\
            \bottomrule
        \end{tabular}
        \caption{Sweep over the number of clusters $K$ for second-level topics on XUB.}
        \label{tab:k_sweep}
    \end{table}

    Topic coherence is evaluated using standard metrics provided by the Gensim library, including NPMI and UMass scores.
    Topic diversity is measured as the proportion of distinct topic phrases relative to the total number of topic assignments.

    Fixation metrics are computed over sliding time windows of 7 days unless otherwise specified, capturing short-term behavioral trends in user engagement.
    We set $\alpha{=}\beta{=}\gamma{=}\tfrac{1}{3}$ to assign equal contribution to diversity, dominance, and recurrence in the absence of validated ground-truth weighting schemes; this neutral setting avoids privileging any single dimension and keeps the composite index interpretable.

    All experiments are conducted using PyTorch and Hugging Face Transformers on a machine equipped with an NVIDIA A100 GPU (80GB). The minimal hardware requirement for running the code is two NVIDIA RTX 3090 GPUs (24GB each).

    \section{Topic Clustering Performance}\label{sec:topic-clustering-performance}
    \begin{figure}[t]
        \centering
        \includegraphics[width=0.5\textwidth]{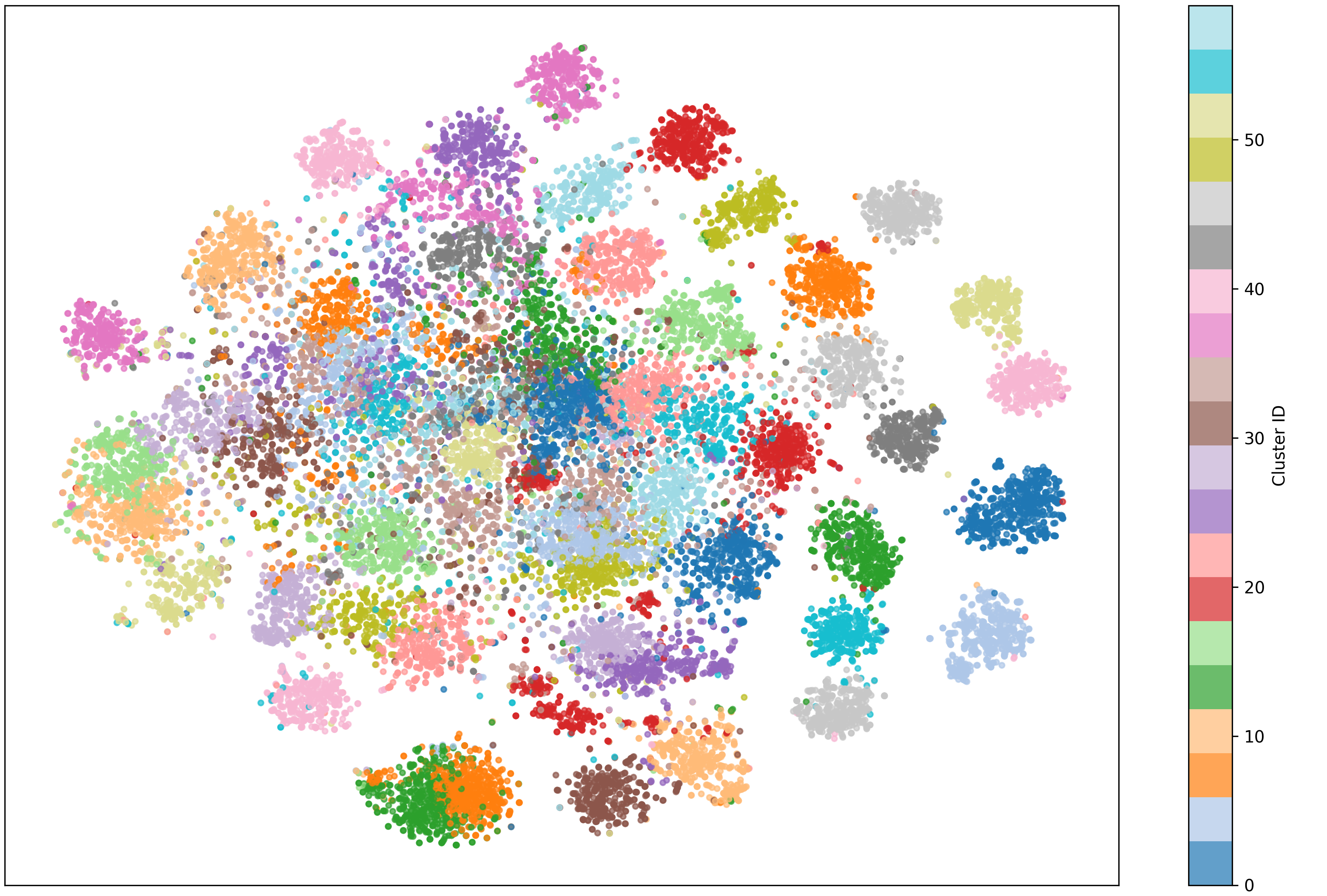}
        \caption{Top 60 clustering results of first-level topics.}
        \label{fig:cluster_viusal}
    \end{figure}

    We perform clustering on the first-level labels of the multimodal data, and the visualization results are shown in Figure~\ref{fig:cluster_viusal}.
    Each point in the figure represents a multimodal sample consisting of a document and a video, with different colors indicating different clustering categories, totaling 60 classes.
    Overall, the clustering results are promising, as clear cluster structures are formed, suggesting that the model effectively distinguishes different topics at the semantic level.
    Some clusters are compact and well-separated, indicating strong internal consistency, while others are more dispersed, reflecting greater diversity within those topics.
    In the central area, overlapping colors suggest that some samples may involve multiple topics or that there are semantic correlations between certain topics, making them harder to separate in the reduced-dimensional space.
    In terms of specific clusters, isolated clusters located in the bottom right and bottom left corners indicate topics that are significantly different from the rest.
    The varying sizes of clusters also reflect the imbalance in topic distribution, where larger clusters correspond to mainstream topics, and smaller ones may represent more niche or specific content.

    \section{Human Annotation Protocol}\label{sec:human-annotation}
    This appendix documents the data collection and human annotation protocol in sufficient detail to enable replication. It describes sampling and preparation of materials, annotator recruitment and training, instructions and decision rules, quality control, privacy safeguards, and output specifications. No experimental results are reported here.

    \subsection{Data Sampling and Preparation}
    We sampled \textbf{30} anonymized users from XUB for human labeling. For each user we prepared an \emph{annotation packet} containing:
    \begin{itemize}
        \item \textbf{Interaction timeline}: a chronological list of viewed items with timestamps, each mapped to second-level topic clusters.
        \item \textbf{Daily topic shares}: stacked bars showing per-day distribution over topic clusters (top~10 displayed; tail aggregated).
        \item \textbf{Word cloud}: top keywords aggregated from first-level topics belonging to the dominant clusters.
    \end{itemize}
    All packets were stripped of personally identifiable information (PII): usernames, profile photos, raw post texts, media, and exact timestamps were removed or coarsened (day-level) prior to handoff to annotators.

    \subsection{Annotator Recruitment and Training}
    We recruited three trained annotators with prior experience in content analysis.
    Before formal labeling, annotators completed a \textbf{calibration round} on five held-out users, followed by a 30-minute group discussion to align interpretations of the guidelines.
    Disagreements were documented to refine the instructions below.

    \subsection{Task Definition}
    Annotators assign a user-level label based on the 30-day packet:
    \begin{itemize}
        \item \textbf{Fixated}: the user exhibits sustained, repetitive engagement concentrated in a narrow topical domain.
        \item \textbf{Not fixated}: the user maintains broad or shifting interests without persistent, narrow focus.
    \end{itemize}
    Annotators should rely on the packet visualizations rather than any external information. Labels reflect the \emph{entire} 30-day window, not isolated days.

    \subsection{Decision Rules and Cues}
    Annotators apply the following operational cues. None is sufficient alone; decisions should weigh them holistically.
    \begin{description}
        \item[Topical concentration (Dominance).] One or two clusters consistently occupy a large share across many days; the head cluster remains dominant with limited rotation.
        \item[Topical breadth (Diversity).] Low variety in active clusters; repeated recurrence of the same few clusters; new clusters rarely appear or vanish quickly.
        \item[Temporal persistence (Recurrence).] Regular revisit rhythms to the same cluster (e.g., long streaks with short inter-visit gaps), indicating routine engagement rather than sporadic bursts.
    \end{description}

    \subsection{Annotation Interface and Materials}
    Annotators worked in a browser-based interface displaying the packet panes (timeline, daily shares, word cloud) and a single-choice control for the user-level label.
    A free-text rationale field (1--3 sentences) was required to summarize the decisive cues (e.g., ``dominant cooking cluster with 18-day streak'').
    Average time per user was approximately 5 minutes during the formal round.

\end{document}